\documentclass[aps,prd,letterpaper,twocolumn,groupedaddress,showpacs]{revtex4}
\usepackage{amsmath,amssymb,graphicx}
\def\bea{\begin{eqnarray}}
\def\eea{\end{eqnarray}}

\def\beq{\begin{equation}}
\def\eeq{\end{equation}}
\def\beqa{\begin{eqnarray}}
\def\eeqa{\end{eqnarray}}

\newbox\pippobox

\begin{document}

\title{Cosmological Coincidence without Fine Tuning}

\author{Joohan Lee}
\email{joohan90@hotmail.com}
\altaffiliation[on leave of absence from]{~Department of Physics, University of Seoul, Seoul 130-743 Korea}
\affiliation{Department of Physics, Astronomy and Geosciences, Towson University, Towson, MD 21210, USA}

\author{Tae Hoon Lee}\email{thlee@ssu.ac.kr}
\affiliation{Department of Physics, Soongsil University, Seoul 156-743 Korea}

\author{Phillial Oh}
\email{ploh@newton.skku.ac.kr}\affiliation{Department of Physics and Institute of Basic Science, Sungkyunkwan University, Suwon 440-746 Korea}

\author{James M. Overduin}
\email{joverduin@towson.edu}
\altaffiliation[also at]{~Department of Physics and Astronomy, Johns Hopkins University, Baltimore, MD 21218, USA}
\affiliation{Department of Physics, Astronomy and Geosciences, Towson University, Towson, MD 21210, USA}

\date{\today}

\begin{abstract}
We present a simple cosmological model in which a single, non-minimally coupled scalar field with a quartic potential is responsible for both inflation at early times and acceleration at late times.
Little or no fine tuning is needed to explain why the present density of dark energy is comparable to that of pressureless matter.
Dark energy is identified with the potential of the scalar field, which is sourced by the trace of the energy-momentum tensor.
This becomes significant when matter has decoupled from radiation and become fully non-relativistic, so that $\phi\propto\rho_m^{1/3}\propto\rho_{m,0}^{1/3}(a_0/a)\sim(10^{-120})^{1/3}(10^{10})\sim10^{-30}$ and $V\sim\phi^4\sim10^{-120}$ in Planck units, as observed.
\end{abstract}

\maketitle

\section{Introduction} \label{secI}

Both early inflation and the recent accelerated expansion of our universe can be explained using scalar fields with very flat potentials, raising hopes that the same scalar might be responsible for both.
However, the energy densities involved differ vastly, by some 120 orders of magnitude.
It is a challenge to build a model which can interpolate smoothly between the two regimes without fine-tuning, and without spoiling the success of the standard big bang model. 
There have been several attempts to construct a unified theory of inflation and dark energy  \cite{Ford,Spok,JnP,PnV,FKL,PnR,Kag,DnV, Wet,HMSS}, in some cases also including dark energy and/or the Higgs field in the same framework \cite{LnU,Sha}.
A recent review of what has come to be known as ``quintessential inflation'' has been given in Ref.~\cite{Hos}.

In this paper we present one such model based on general relativity and ideas from induced-gravity (IG) theory \cite{Sak,Zee}, in which gravitation is regarded as arising from quantum processes within ordinary field theory. 
One-loop effects in scalar-field theories produce a term corresponding to the non-minimal coupling of the scalar field to the scalar curvature of the spacetime.
In IG theory this term is taken as the action for gravity. 
Here, we retain the standard Einstein-Hilbert action and regard the IG coupling term as an additional part of the action. 
However, we insist that no dimensionful parameters other than the Planck mass be introduced. 
Sec.~\ref{secII} describes the model and discusses the general behavior of the scalar field in the  cosmological context. 
In the Einstein frame the shape of the potential is such that the scalar field initially rolls slowly with energy of order one in Planck units, and later decays rapidly into relativistic particles, reheating the universe. 
After reheating is over the scalar field, having lost all of its energy, is assumed to stay near the ground state.

As massive particle species such as baryons or WIMPs drop out of equilibrium and become non-relativistic, the scalar field begins to grow until balanced by a restoring force due to the potential gradient. 
A straightforward numerical estimate shows that the potential energy stored in the scalar field at this time is within a few orders of magnitude of the present-day density of pressureless matter.
The coincidence problem is thus solved if the value of the scalar field generated at that time remained almost constant until now.
We give an argument showing that this is indeed possible. 
These aspects of the model are discussed in Sec.~\ref{secIII}.
In Sec.~\ref{secIV} we compare our model to others in the literature, and outline some of the ways in which it can be tested in future work.

\section{The model} \label{secII}

Our Lagrangian resembles that of pure IG theory \cite{Zee} but also retains a role for the Planck mass (as in standard general relativity):
\beqa
L & = & \sqrt{-g}\left[\vphantom{\frac12}(c_1M_P^2+c_2M_P\phi +c_3 \phi^2)R \right. \nonumber \\
& & \left. -{1\over2}K(\phi)g^{\mu\nu}\partial_\mu\phi\partial_\nu\phi-{\lambda\over4}\phi^4\right].
\label{action1}
\eeqa
We have included $K(\phi)$ in the kinetic term for generality.
A non-canonical form for kinetic energy is characteristic of k-essence cosmology \cite{Kess, Kess2}, and our choice can be regarded as its low energy limit.
In the context of a single-field model, it can be transformed into the canonical form by a redefinition of the scalar field.

Note that the $c_2$ term breaks the symmetry $\phi\rightarrow -\phi$.
We rewrite Eq.~(\ref{action1}) as
\beqa
L & = & \sqrt{-g}\left\{{\vphantom{\frac12}1\over2}M_P^2\left[(\phi/M_P-\alpha)^2+\beta^2\right]R \right. \nonumber \\
& & \left. -{1\over2}K(\phi)g^{\mu\nu}\partial_\mu\phi\partial_\nu\phi-{\lambda\over4}\phi^4\right\},
\label{action2}
\eeqa
ensuring that the coefficient in front of the Ricci scalar is always positive.
Without loss of generality we also assume that $\alpha$ is positive. 

When the field is large compared to the Planck mass one can show that the scale symmetry is restored.
This means that the field can be almost stationary at any large value.
For small values of the field, on the other hand, the scaling symmetry is broken and $\phi=0$ becomes the unique stable equilibrium configuration.
We wish to recover standard general relativity in the vicinity of the vacuum, so we require
$\alpha^2+\beta^2=1 $.

\begin{figure}
\begin{center}
\includegraphics[width=\columnwidth]{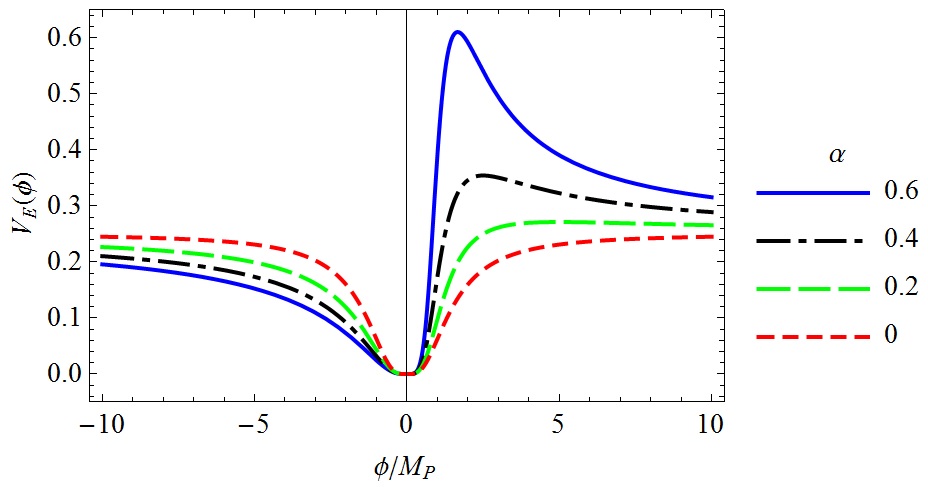}
\caption{The effective potential $V_E(\phi)$ in the Einstein frame for several values of $\alpha$ (with $\phi$ in Planck units and $\lambda=1$).}
\label{fig1}
\end{center}
\end{figure}

The dynamics of the the scalar field can be described more easily in the Einstein frame.
For this purpose we make a conformal transformation $g_{\mu\nu} \rightarrow \chi^{-2}g_{\mu\nu}$ where
\beq
\chi^2(\phi) \equiv (\phi/M_P-\alpha)^2+\beta^2.
\label{chiDefn}
\eeq
Then the Lagrangian~(\ref{action2}) becomes
\beq
L = \sqrt{-g}\left[{1\over2}M_P^2 R-{1\over2}K_E(\phi)g^{\mu\nu}\partial_\mu\phi\partial_\nu\phi-V_E(\phi)\right],
\eeq
where
\beqa
K_E(\phi) & \equiv & {K+6M_P^2{\chi^\prime}^2\over\chi^2} , \\
V_E(\phi) & \equiv & {\lambda\over4}\phi^4\chi^{-4}(\phi).
\label{Vdefn}
\eeqa
Apart from a non-canonical kinetic term, this is just the Lagrangian for a scalar field with minimal coupling to Einstein gravity and a dilatonic coupling to matter.
The form of the potential is illustrated in Fig.~\ref{fig1}.
For large negative values of $\phi$ the potential becomes flat and asymptotically approaches $\tfrac{1}{4}\lambda M_P^4$, which is of order one in Planck units.
For smaller negative values the potential drops rapidly to zero.
Near this minimum, the potential is approximately quartic.
For positive values of $\phi$ the potential has a bump around $\phi=\alpha^{-1}M_P$ whose height is $\beta^{-2}$ times the asymptotic value.

Assuming that the scalar field starts with a large negative value, we can divide its evolution into three regimes: (i)~$\phi\ll-M_P$, (ii)~$\phi\sim -M_P$, and (iii)~$-M_P\ll\phi<0$.
Since the potential is almost flat for large negative $\phi$, its potential energy will quickly dominate over kinetic and other forms of energy and the universe will undergo almost exponential expansion during regime~(i), representing inflation.
Similar ideas have been discussed in the context of ``induced-gravity inflation'' \cite{IGinflation}.

After sufficient inflation when the scalar field is of the order of the Planck mass, it will rapidly decay due to the steep slope of the potential, reheating the universe by transferring its energy to other forms of matter and ushering in the radiation-dominated era.
The details of this and the resulting state of the scalar field are necessarily model-dependent  \cite{Pert,Irh,ABCM}.
In the scheme of Ref.~\cite{Irh}, the scalar field decays almost completely, while in other cases it comes into thermal equilibrium with standard-model fields (as in tracking-type models \cite{ZWS}).
In this work we will simply assume that after reheating the scalar field settles very close to the the minimum of the potential, and concentrate on the behavior of $\phi$ in regime~(iii) in order to discuss the cosmological coincidence problem. 

\section{A mechanism for cosmological coincidence} \label{secIII}
 
The scalar field in this model is sourced by the trace of the energy-momentum tensor, $\rho-3p$.
Thus it begins to grow again as the universe transitions from radiation to matter dominance.
This can be seen from the field equations, which read as follows (for a flat homogeneous and isotropic metric):
\beqa
& & M_P^2 3H^2 = {1\over2}K_E\dot{\phi}^2+V_E+\rho , 
\label{fieldEq1} \\
& & \hspace{-7mm} K_E(\ddot{\phi}+3H\dot{\phi})+{1\over2}{dK_E\over d\phi}\dot{\phi}^2 = -{dV_E\over d\phi}+{\chi^\prime\over\chi}(\rho-3p),
\label{fieldEq2}
\eeqa
where primes denote derivatives with respect to $\phi$, and $\rho$ and $p$ refer to the energy density and pressure of additional matter in the Einstein frame, which is not conserved due to conformal coupling of the scalar field.
In principle this conformal coupling can lead to severe experimental constraints.
Here, however, its effects are undetectable during the epoch relevant to our main discussion because the value of the scalar field is extremely small, as shown below.

In what follows, we make a specific ansatz for the kinetic energy factor in the neighborhood of $\phi=0$, 
\beq
K(\phi) \equiv \xi{M_P^2\over\phi^2},
\label{kineticTerm}
\eeq
where the Planck mass is inserted on dimensional grounds.
This form of kinetic energy factor is introduced to slow the scalar as it rolls toward zero.
In fact, it never actually reaches zero or becomes positive.
This is best seen using a new field variable $\sigma\equiv-M_P\sqrt{\xi}\log(-\phi/M_P)$, in terms of which the model goes over (for small $\phi$) to one with a minimal coupling to gravity, conformal coupling to matter, and an exponential potential, $V(\sigma)\sim e^{-4\sigma}$.
The field value $\phi=0$ corresponds to $\sigma=\infty$.

In the vicinity of the minimum of the potential, Eqs.~(\ref{fieldEq1}) and (\ref{fieldEq2}) can then be approximated as follows, keeping only the leading terms:
\beqa
& & M_P^2 3H^2={\xi\over2}M_P^2\left({\dot{\phi}\over\phi}\right)^2+{\lambda\over4}\phi^4+\rho \label{eom1}, \\
& & \xi{M_P^2\over\phi}{d\over dt}\left({\dot{\phi}\over\phi}\right)=-\xi{M_P^2\over\phi}3H\left({\dot{\phi}\over\phi}\right)-{d\over d\phi}({\lambda\over4}\phi^4) \nonumber \\
& & \hspace{3cm} -{\alpha\over M_P}(\rho-3p). \label{eom2}
\eeqa
A few remarks are in order.
First, at early times when all matter was relativistic the last term on the right-hand side of Eq.~(\ref{eom2}) did not directly contribute to the the expectation value of the scalar field.
Second, the friction term becomes large as $\phi$ approaches zero, so that the scalar field remains near this minimum for a long time.
However, as the universe cools and matter decouples from radiation, the scalar field begins to feel a force due to the growing density of non-relativistic matter.
It then begins to be pulled back toward negative values.
We assume that the magnitude of the scalar field reaches a maximum value (denoted $\phi_{\ast}$) when the potential gradient balances the force due to matter.
At this point, its velocity is zero, the acceleration term is small, and matter is fully non-relativistic.
(The same results follow as long as the ratio of the acceleration term to friction approaches a constant, as in Ref.~\cite{CDS}.)
 
The value of $\phi_{\ast}$ can be estimated using Eq.~(\ref{eom2}):
\beq
-\lambda\phi_{\ast}^3={\alpha\over M_P}\rho_{m,\ast} = {\alpha\over M_P}\rho_{m,0}\left({a_0\over a_{\ast}}\right)^3,
\eeq
where $\rho_m$ denotes pressureless matter and the subscripts $0$ and $\ast$ denote respectively the present time and the time when the above balance is reached.
The potential energy of the scalar field at this time is  $V_{\ast}=(\lambda/4)\phi_{\ast}^4$ from Eq.~(\ref{Vdefn}), where $\chi(\phi)\approx 1$ from Eq.~(\ref{chiDefn}) since $\phi\ll M_P$.
The {\it ratio} of this energy relative to the present density of matter is then given by
\beq \frac{V_{\ast}}{\rho_{m,0}}={\lambda\over4}\left({\alpha\over\lambda}\right)^{4\over3}\left({\rho_{m,0}\over M_P^4}\right)^{1\over3}\left({a_0\over a_{\ast}}\right)^4.
\label{wow}
\eeq
To estimate this quantity, we note that
$\rho_{m,0}/M_P^4=24\pi\hbar G\Omega_{m}H_0^2/c^5=3\times10^{-121}$ in physical units (with $M_P$ the reduced Planck mass),
where we have used WMAP values for $\Omega_{m}$ and $H_0$ \cite{WMAP}.
If we restrict our attention to massive particle species originally in thermal equilibrium with radiation, then $a_0/a_{\ast}\approx 3(T_{\ast}/T_0)$, where the CMB temperature $kT_0=0.2$~meV and $T_{\ast}$ is determined by the particle mass $m$.
(As these particles freeze out of equilibrium, they deposit energy into the cosmic plasma so that product $aT$ does not remain constant \cite{Dodelson}.)
Particles begin to freeze out when $kT\sim mc^2$, but become fully non-relativistic only when $kT\sim mc^2/30$ (\cite{KolbTurner}, Fig.~5.1).
We thus take
\beq
\frac{a_0}{a_{\ast}}\approx 3\frac{T_{\ast}}{T_0}\sim\frac{1}{10}\frac{mc^2}{kT_0}=4\times10^{11}
\label{a0aStar}
\eeq
for baryons in particular.
If, however, non-relativistic matter is dominated by thermal cold dark-matter relics (which also dropped out of equilibrium with radiation), then the scalar field would be sourced primarily by them.
There have been recent hints from multiple direct detection experiments that the CDM might consist of supersymmetric WIMPs with $m\sim10$~GeV$/c^2$ \cite{DelNobile}; in this case Eq.~(\ref{a0aStar}) gives instead $a_0/a_{\ast}\sim4\times10^{12}$.

Another strong CDM candidate is the axion, with $m\sim10^{-5}$~GeV$/c^2$.
These particles, however, are non-thermal (they were never in equilibrium with radiation).
They are ``semi-relativistic'' at birth near the QCD phase transition  (i.e., when $kT\sim\Lambda_{\mbox{\tiny QCD}}\approx150$~MeV) but rapidly settle down to form a zero-momentum condensate as $kT\ll\Lambda_{\mbox{\tiny QCD}}$ (\cite{KolbTurner}, Ch.~10).
(There are also thermal axion models, but these have been ruled out by astrophysical and cosmological bounds \cite{OW}.)
Following the same reasoning as above, we thus take for these particles
\beq
\frac{a_0}{a_{\ast}}\sim\frac{1}{10}\frac{\Lambda_{\mbox{\tiny QCD}}}{kT_0}=6\times10^{10}.
\label{a0aAxion}
\eeq
Eq.~(\ref{wow}) with these numerical results implies
\beq
\frac{V_{\ast}}{\rho_{m,0}}\sim\lambda\left(\frac{\alpha}{\lambda}\right)^{4\over3}\times\left\{ \begin{array}{ll}
200 & \mbox{ (axions) } \\
4\times10^5 & \mbox{ (baryons) } \\
4\times10^9 & \mbox{ (WIMPs) }.
\end{array} \right.
\label{tune}
\eeq
In order to ensure that the potential energy of the scalar field has remained constant or decayed only slowly since $t_{\ast}$
consider that during this period the potential gradient is balanced by the Hubble friction so that
\beq
\xi{M_P^2\over\phi^2}3H\dot{\phi}=-{d\over d\phi}({\lambda\over4}\phi^4)=-\lambda\phi^3 .
\eeq
From this we can estimate the ratio of kinetic to potential energy as
\beqa
{(1/2) \xi M_P^2({{\dot\phi}/\phi})^2\over (\lambda/4)\phi^4} & = & { \lambda^2\phi^8/18\xi M_P^2 H^2\over (\lambda /4)\phi^4} \nonumber \\
& = & {8\over 3\xi} {(\lambda/4)\phi^4\over3H^2 M_P^2} \ll 1 .
\eeqa
This ratio remains small throughout the subsequent evolution of the scalar field, as required.

It is thus natural to identify $V_{\ast}$ with the present-day dark-energy density $\rho_{\Lambda,0}$.
Eq.~(\ref{tune}) then shows that little or no fine tuning of the model parameters $\alpha$ and $\lambda$ is required to explain the cosmological coincidence that $(\rho_{\Lambda,0}/\rho_{m,0})_{\mbox{\tiny obsd}}\approx3$.
Taking $\alpha\sim1/\lambda$ for convenience, we find that $\alpha\sim10^{-1}$, $10^{-3}$ and $10^{-6}$, assuming matter dominated by axions, baryons and WIMPs respectively.
(These values could be even closer to unity if the potential energy of the scalar field has in fact decreased slightly since $t_{\ast}$.)
The implied degree of tuning is negligible for axions but possibly significant for WIMPs, suggesting that in the context of this model at least, axions are to be preferred as CDM candidates.


\section{Summary and discussion} \label{secIV}

We have constructed a cosmological model without small parameters in which a single scalar field may be responsible both for inflation at early times and acceleration at late ones.
We have concentrated here on the potential of this model to provide a natural explanation for the present density of the dark energy.
Little or no fine tuning appears to be required in either the model parameters or initial conditions, possibly providing a new way to address the coincidence problem.

The main argument can be summarized as follows.
As the universe cools, a growing fraction of its relativistic matter content gradually becomes non-relativistic.
This pressureless matter exerts a force on the scalar field (located close to zero at this time), pulling it back toward negative values until balanced by the potential gradient.
By estimating the temperature at which matter (dominated by baryons or cold dark matter in the form of WIMPs or axions) became fully non-relativistic, we have shown that the potential energy associated with the scalar field at this time is of order $10^{-120}$ in Planck units.
This energy changes very little during the subsequent evolution of the universe, so it can explain the density of dark energy measured today.

Our results suggest a new way of looking at dark energy.
This picture is different from that in quintessence theory \cite{Quint}, where a scalar field slowly rolls indefinitely toward zero, eventually becoming dominant over matter.
In these models, some fine tuning in both model parameters and initial conditions is generally required to explain why the dark-energy density is comparable to that of matter at the present time. 
The underlying mechanism here is similar in some respects to one in Ref.~\cite{Kam}, where non-relativistic matter also ``pulls'' a scalar field out of the ground state at late times.
However, our model differs in motivation, in the behavior of the scalar field, and in the degree of tuning required.
The scalar field in our model may be responsible for inflation as well as acceleration.
(It is difficult to  satisfy the slow-roll conditions for inflation with a canonical kinetic term for the scalar field, as in Ref.~\cite{Kam}.)
The potential in our model is everywhere smooth and finite (in the Einstein frame), while that in Ref.~\cite{Kam} becomes singular when the field reaches zero. 
The detailed behavior of the scalar field around the ground state is also different, which affects the dynamics both during inflation and at late times. 
In Ref.~\cite{Kam}, the growth of the scalar field at late times is tied to the epoch at which matter begins to dominate over radiation, rather than the epoch at which it becomes non-relativistic.
Because of the sensitive dependence of the value of the scalar potential on this choice of epoch, the two models lead to quite different results.

It would be of great interest to test models of this kind via the power spectrum of large-scale structure \cite{ZWS} or the magnitude-redshift relation for Type~Ia supernovae.
The latter requires solutions of the field equations~(\ref{eom1}) and (\ref{eom2}) for the scale factor $a$ over the redshift range $0\leqslant z\lesssim2$.
We hope to report on constraints of this kind in the near future \cite{Nathan}.

To test the inflationary aspects of the model, a numerical phase-space analysis of the dynamical system would be valuable, varying the scalar field parameters and evaluating the stability of solutions with respect to the initial conditions.
Such a test would however depend strongly on specific assumptions about the nature of the reheating process, as well as the identity of dark matter.

Other tests could come from the spectral index and tensor-to-scalar ratio of CMB fluctuations \cite{HosCMB}.
We emphasize that the $\lambda\phi^4$-term in the potential does not mean that inflation is of the standard chaotic type, where $\lambda$ is constrained by current CMB data to be less than $\sim10^{-13}$ \cite{CMB}.
As discussed above, the $\phi^4$ part of the action is relevant only for small values of the scalar field, whereas inflation takes place when $-\phi$ is large and the potential is dominated by the other terms in Eq.~(\ref{action1}) and is nearly flat.
In this sense our model more closely resembles ``new'' inflation, and the strong observational constraints that have been obtained on $\phi^4$-type chaotic inflation models in particular do not apply.

\acknowledgments
JL thanks Towson University for hospitality as a sabbatical visitor.
JL's work was supported by 2013 Research Fund of the University of Seoul.
THL was supported by the Basic Science Research Program through the National Research Foundation of Korea (NRF) funded by Ministry of Education, Science and Technology (NRF-2010-0012692). PO was supported by the National Research Foundation of Korea (NRF) grant funded by the Korea government (MEST) through the Center for Quantum Spacetime (CQUeST) of Sogang University with grant number 2005-0049409 and by the BSRP through the National Research Foundation of Korea funded by the MEST (2010-0021996).

\end{document}